\documentclass[aps,prl,superscriptaddress,showpacs,twocolumn]{revtex4}
\usepackage{epsfig}
\begin{document}
\title{Discrete charging of a quantum dot strongly coupled to external leads}
\author{Richard Berkovits}
\affiliation{The Minerva Center, Department of Physics,
    Bar-Ilan University, Ramat-Gan 52900, Israel}
\author{Felix von Oppen}
\affiliation{Institut f\"ur Theoretische Physik, Freie Universit\"at Berlin, Arnimallee
14, 14195 Berlin, Germany}
\author{Jan W. Kantelhardt}
\affiliation{Institut f\"ur Theoretische Physik III, Justus-Liebig-Universit\"at Giessen,
35392 Giessen, Germany}
\date{May 18, 2004, version 8.0}
\begin{abstract}
We examine a quantum dot with $N_{\rm dot}$ levels which is 
strongly coupled to
leads for varying
number of channels $N$ in the leads. It is shown both analytically and numerically that 
for strong couplings between the dot
and the leads, at least $N_{\rm dot}-N$ bound states 
(akin to subradiant states in optics) remain on the dot.
These bound states exhibit discrete charging and, for a significant range of
charging energies,
strong Coulomb blockade behavior as function of the chemical
potential. The physics changes for large charging energy where the same (superradiant) state 
is repeatedly charged. 
\end{abstract}

\maketitle

It is well known that the number of electrons in a weakly coupled quantum dot changes
discretely as function of the chemical potential. This phenomenon is the basis for the
application of such dots as single electron transistors \cite{kastner92}. When the 
coupling to the external leads is weak, it may be treated as a perturbation and results
in a broadening $\Gamma=  \pi N \nu |V|^2$
of the states of the uncoupled dot. ($\nu$ is the
density of states (DOS) in a lead, $N$ the number of leads,
and $V$ is the overlap matrix element between a state in
the dot and a typical state of the leads.) Usually, one expects the discrete features of
the dot to be lost once $\Gamma$ is larger than the typical level spacing $\Delta$ in the
dot. This corresponds to the requirement that the dimensionless conductance through the
dot, $g_{\rm dot}=\Gamma/\Delta$, should be larger than one.

Electron-electron interaction in the dot results in the Coulomb blockade phenomenon
\cite{kastner92}. Within the "orthodox model" \cite{alhassid00}, the chemical potential
change needed to add an additional electron to a weakly coupled dot is no longer $\Delta$
but rather $\Delta+e^2/C$, where $C$ is the capacitance of the dot. For stronger coupling
to the leads, suppression of the Coulomb blockade is predicted \cite{aleiner02}, and once
$\Gamma \sim \Delta$ only a weak remnant of the discreteness of the dot is expected
\cite{matveev95}.

In this picture the only relevant condition for the the appearance of discrete
features in the dot is $g_{\rm dot}<1$. Nevertheless, one may extrapolate from the Dicke
effect \cite{dicke53,shah98}  that a quantum dot strongly coupled to a
lead will also show sharp resonances. In an extreme strong coupling limit, we 
can think of the dot levels as degenerate, analogous to the identical resonances of the atoms
in the Dicke effect. These degenerate levels are coupled via the continuum of lead states
which is akin to the coupling of the atoms by the radiation field. 
Indeed, a resonance in a two orbital dot strongly
coupled to two leads was seen in Ref. \cite{shah94}. 
The tunneling DOS of a
non-interacting two-orbital dot strongly coupled to a single lead tends towards a
delta-like peak when $g_{\rm dot} \rightarrow \infty$ \cite{konig98}. It was later shown
\cite{montambaux99} that when a dot of $N_{\rm dot}$ states is strongly connected to a
single lead, $N_{\rm dot}-1$ delta-like peaks in the tunneling DOS remain
for $g_{\rm dot} \rightarrow \infty$. The case of two leads connected to a dot with
$N_{\rm dot}=2$ shows interesting dependence on the overlap matrix element $V_{k,i}$
(i.e., the $i$-th level overlap with $k$-th lead) \cite{silva02}. For identical matrix
elements one sharp peak in the local density of states remains, while if one of the four
matrix elements has a different sign, no such features are observed. 

In this letter we show that in the limit $g_{\rm dot} \rightarrow \infty$, the relevant
parameter in determining the number of bound states in the dot for generic dot-lead coupling 
is the number of channels $N$ of the leads, or equivalently, the dimensionless conductance 
$g$ of the leads. We demonstrate that for
$N$ channels coupled to a dot, $N_{\rm dot}-N$ states remain
bound to the dot, except when the coupling matrix elements between leads and dot are
independent of either dot level or channel index.
For the latter cases, $N_{\rm dot}-1$ states remain bound to the dot. 
Using numerical density-matrix renormalization group
(DMRG) as well as analytical arguments, we show that these bound states exhibit 
discrete charging as well as Coulomb blockade except for very large charging 
energies. Thus, in order to wash out all discrete features of
a dot in the limit of strong coupling, one or more leads of total dimensionless
conductance $g>N_{\rm dot}$ must be connected. With increasing coupling strength, a dot
coupled to leads will evolve from $N_{\rm dot}$ bound states at very weak coupling, to no
bound states at intermediate coupling \cite{aleiner02} and finally to $N_{\rm dot}-g$
bound states at strong coupling. Criteria for the different regimes and experimental
realizations will be discussed. It is interesting to note that the physics discussed 
here for quantum dots is also closely related to the concept of doorway states in 
nuclear physics \cite{Mahaux}.

We consider a dot-lead system described by the Hamiltonian
\begin{equation}
H = H_{\rm dot} + \sum_{k=1}^N \left( H^k_{Lead} + H^k_{Coupling} \right).
\label{hamiltonian}
\end{equation}
Here, the dot is represented by the Hamiltonian
\begin{equation}
H_{\rm dot}=\sum_{i=1}^{N_{\rm dot}} (\epsilon_i -\mu) a_i^{\dag} a_i + U \sum_{i>j}^{N_{\rm dot}}
a_i^{\dag} a_i a_j^{\dag} a_j \label{hqd}
\end{equation}
in terms of the creation operators $a_i^{\dag}$ of an electron in the $i$-th
single-particle eigenstate of the dot with energy $\epsilon_i$, charging energy $U=e^2/C$
and chemical potential $\mu$. For disordered dots its eigenstates and eigenvalues are
usually obtained from a random matrix ensemble. The Hamiltonian of the $k$-th lead reads
\begin{equation}
H^k_{Lead}= \mu \sum_{j=1}^{\infty}c_j^{k \dag} c^k_{j} - t \sum_{j=1}^{\infty} c_j^{k
\dag} c^k_{j+1} + h.c.,\label{hlead}
\end{equation}
where $c_j^{k \dag}$ is the creation operator of an electron on the $j$-th site of the
$k$-th 1D lead, and $t$ is the hopping matrix element in the lead.  The coupling between
the dot and the lead is contained in
\begin{equation}
H^k_{Coupling}= \sum_{i=1}^{N_{\rm dot}} V_{k,i} a_i^{\dag} c^k_{1} + h.c.,\label{hdl}
\end{equation}
where the dot is assumed to be attached to the edge of the lead, and the coupling
amplitude between the $i$-th orbital in the dot and the $k$-th lead is given by $V_{k,i}$.
The $N$
one-dimensional leads may also be connected by transverse hopping
$-t \sum_{k=1}^{N-1} \sum_{j=1}^{\infty} c_j^{k \dag} c^{k+1}_{j} + h.c.$ in order
to turn them into a quasi-one dimensional lead with $N$ channels.

We begin with exact-diagonalization results for a 
non-interacting ($U=0$) dot coupled to an external quasi-1D lead of varying width
(i.e., number of channels $N$). 
Exact diagonalization can only
treat finite systems, and therefore cannot deal with infinite leads. Nevertheless, as
long as the level broadening in the dot is much larger than the level spacing in the
lead, the description of the system is accurate. 
Diagonalizing of the Hamiltonian $H$ gives
the eigenvalues $\varepsilon_m$ and eigenvectors $|m \rangle$ of the dot-lead system. The
number of electrons on the dot at a given chemical potential is $n =
\sum_{m=1}^{\varepsilon_m<0} \sum_{i=1}^{N_{\rm dot}} | \langle m | a_i^{\dag} a_i |m
\rangle |^2$. 

\begin{figure}\centering
\epsfxsize8cm\epsfbox{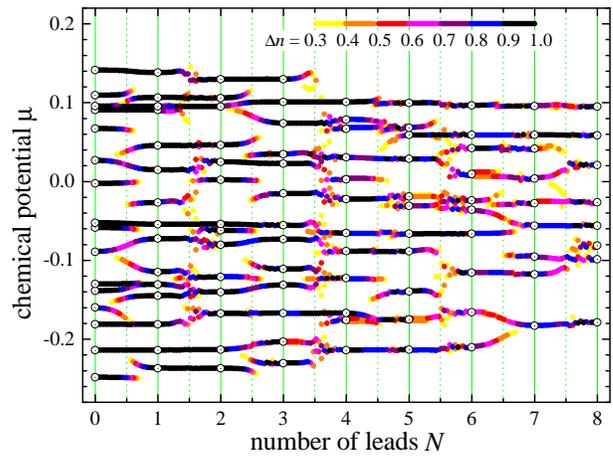} \vskip .3truecm \caption{ The
increases $\Delta n(\mu,N)$ in the occupation of a quantum dot
with $N_{\rm dot}=16$ orbitals as function of the chemical
potential $\mu$ and the width $N$ of the quasi-one-dimensional lead
(length $M=100$) connected to the dot. Non-integer numbers of $N$
correspond to having the connection of the last channel to the dot and 
its neighboring channel
logarithmically increased (see text). The increase in the
occupation is color coded as indicated in the figure. The symbols
correspond to the derivative in the limit of $V \rightarrow
\infty$} \label{fig1}
\end{figure}

In Fig.\ \ref{fig1} we present the discrete increases $\Delta n(\mu,N) =
\int_{\mu-0.001t}^{\mu+0.001t} (dn/d\mu') \, d\mu'$ in the occupation of a disordered
quantum dot with $N_{\rm dot}=16$ orbitals (generated by a random matrix with a Gaussian
distribution of width $0.1t$) as function of the chemical potential $\mu$ and 
$N$, the number of channels  connected to the dot. In the absence of dot-lead coupling
($N=0$), all increases $\Delta n(\mu,N)>0$ occur at $\mu = \epsilon_m$. The integer points on the
$x$-axis correspond to $N$ channels connected to the dot by couplings
$V_{k,i}$ drawn from a random Gaussian distribution with a zero mean and variance $t$.
The non-integer values of $N$
correspond to increasing logarithmically the values of the coupling of the $N+1$ channel
to the dot and its transverse hopping to the neighboring channel 
up to the full strength $V_{\lfloor N \rfloor+1,i}$ 
and $t$ at integer $N$.
For integer values of $N$ it can be clearly seen that there are
$N_{\rm dot}-N$ jumps in the occupation of the dot corresponding to the same number of
states bound to the dot. As one couples an additional channel to the dot the energies of
these states gradually change, until for some intermediate strength of coupling some of
the states move abruptly, split, or disappear. At stronger couplings (i.e., close to the
next integer $N$) the ordered structure of the states reappears, with one less state than
for the previous value of $N$. Similar behavior is seen when there is no transverse hopping
between the channels (i.e., $N$ independent 1D leads).
If $V_{k,i}=t$ is independent of $i$ or $k$ (i.e., all the
couplings to the same lead or orbital are identical) a loss of a bound state occurs only
when connecting the first lead to the dot. Attaching additional leads does not change the
number of bound states on the dot.

To understand these numerical results we approach the system within the scattering-matrix
formalism. For $N$ propagating channels in the leads, the dot-lead system can be
characterized by the $N\times N$ scattering matrix \cite{zirnbauer}
\begin{equation}
    S = {\bf 1} -2\pi i\nu  V {1\over E - H_{\rm dot} +i\pi\nu V^\dagger V} V^\dagger,
\end{equation}
where $V$ is an $N\times N_{\rm dot}$ matrix describing the coupling of the channels to
the $N_{\rm dot}$ orbitals of the dot. The local DOS of the dot is given by
$dn/ d\mu = {(1/ \pi)} {\rm Im\,tr}[\mu - H_{\rm dot} +i\pi\nu V^\dagger V]^{-1}$.
If, to be specific, we take $H_{\rm dot}$ to be diagonal with random-matrix spectrum, the couplings
$V_{k,i}$ are essentially independent random Gaussian variables of variance $v^2$.
We emphasize, however, that our results are not specific to a random-matrix 
spectrum.

For a weakly-coupled dot, we can compute the S-matrix by first diagonalizing the 
dot Hamiltonian $H_{\rm dot}$.
The broadening of the levels can then be obtained by treating $i\pi\nu V^\dagger V$ in
first-order perturbation theory. To understand the opposite limit of strong coupling, we
first diagonalize $i\pi\nu V^\dagger V$ and subsequently account for $H_{\rm dot}$
perturbatively. We start by writing
$V^\dagger=({\bf v}_1,{\bf v}_2,\ldots,{\bf v}_N)$
where the ${\bf v}_i$ are $N_{\rm dot}$-dimensional vectors. In obvious notation, we can
then write $\pi\nu V^\dagger V=\pi\nu\sum_{i=1}^N |{\bf v}_i\rangle\langle {\bf v}_i|$
which shows that the $N_{\rm dot}\times N_{\rm dot}$ matrix $\pi\nu V^\dagger V$ has at
most rank $N$ and generically only $N$ nonzero eigenvalues $\lambda_{\ell}\sim\pi\nu
N_{\rm dot}v^2$ with $\ell=1,\ldots, N$.

Including $H_{\rm dot}$ perturbatively, we first need to diagonalize $H_{\rm dot}$ in the
$(N_{\rm dot}-N)$-dimensional degenerate subspace of zero eigenvalues. In the limit $N_{\rm
dot}\gg N$, this leads to a random-matrix spectrum of $N_{\rm dot}-N$ real eigenvalues
$e_i$ whose width and level spacing $\Delta$ equal those of the Hamiltonian $H_{\rm dot}$
of the uncoupled dot. Thus, in first-order perturbation theory we find $N_{\rm dot}-N$
infinitely sharp resonances in addition to $N$ imaginary eigenvalues which lead to
an extremely broad background (since $\lambda\sim N_{\rm dot}$) in the local DOS.

In second-order perturbation theory, these resonances acquire a width since $H_{\rm dot}$
couples the sharp resonances to the broad background. The resulting width can be easily
estimated to be
\begin{equation}
   \Delta e_i  \simeq \sum_{\ell=1}^N {|\langle i| H_{\rm dot}|\ell\rangle|^2\over
-i\lambda_{\ell}} \sim i{N\Delta^2\over \pi^2\nu v^2}\sim i{N^2\Delta^2\over \pi \Gamma}.
\end{equation}
Here, we used that $H_{\rm dot}$ is a random-matrix Hamiltonian and defined 
the golden-rule width $\Gamma=\pi\nu N v^2$ of the eigenstates
$\epsilon_i$ of the uncoupled dot. We assume strong
coupling so that $\pi\nu v^2\gg \Delta$ which allows us to neglect the unperturbed energy
$e_i$ in the denominator. When this width remains small compared to the level spacing,
i.e., when $N^2/\pi g_{\rm dot}\ll 1$, we find $N_{\rm dot}-N$ isolated resonances in both
conductance and local DOS even though dot and lead are very strongly
coupled, in agreement with our numerical results.

By an analogous argument one finds only one imaginary eigenvalue and hence $N_{\rm
dot}-1$ sharp resonances in the non-generic cases in which the coupling $V_{k,i}$ is
independent of either channel $k$ or dot level $i$. These resonances have a width of
order $\Delta/\pi g_{\rm dot}$. This width is smaller by a factor $N^2$ compared to the
resonance width in the case of arbitrary dot-lead coupling.

\begin{figure}\centering
\vskip -1.5truecm \epsfxsize6cm\epsfbox{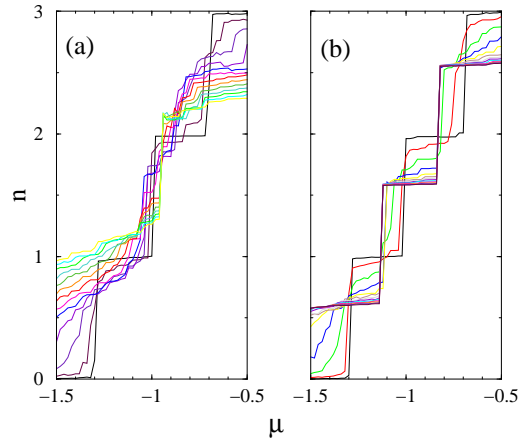}
\vskip .3truecm
\caption{The occupation $n$ as function of the chemical potential
for $N_{\rm dot}=3$ and $N=2$. The dot levels have energies
$\epsilon_1=-1.3t$,$\epsilon_2=-1.2t$, and $\epsilon_3=-1.1t$. The
interaction $U=0.2t$. In (a) the couplings are ,$V_{11}=0.05pt$,
$V_{12}=0.09pt$, $V_{13}=0.01pt$, $V_{21}=0.09pt$,
$V_{22}=0.01pt$, $V_{23}=0.01pt$, where $p=1,2,3 \ldots 12$ for
the different curves, while in (b) all the couplings are
$V=0.05pt$.} \label{fig2}
\end{figure}

We now turn to the influence of the charging energy $U$. We compute the
ground state for an interacting dot attached to several 1D leads 
using an extension of a DMRG method developed for a single 1D
lead connected to a dot \cite{berkovits03} which will be described in detail elsewhere.
As in the single lead case, the essence of the method is similar to the regular DMRG for
1D systems \cite{white93}. The main difference is that in every iteration a site is added
to each of the leads.
Fig.\ \ref{fig2} shows the occupation number $n$ as function of $\mu$ for a $N_{\rm
dot}=3$ dot attached to two leads for different values of the dot-lead coupling.

In Fig.\ \ref{fig2}a, the case of non-identical couplings is presented. The general
behavior seen for the non-interacting case is repeated in the interacting case. For weak
coupling, there are three discrete jumps in the occupation of the dot separated by
$\Delta + U$. As the coupling increases only {\it one} discrete jump remains. Thus, the
interactions in the dot do not eliminate the bound state. Moreover, interactions shift
the position of the remaining jump to higher $\mu$  relative to the noninteracting case.

If the couplings are symmetric to all leads the number of bound states is $N_{\rm dot}-1$ no
matter how many leads are attached. This is illustrated in Fig.\ \ref{fig2}b, where two
discrete jumps remain even for strong coupling. These bound states are separated by a
distance of $\Delta+U$ as one expects from two bound states on an interacting dot.
Similar Coulomb blockade behavior at strong coupling has been recently seen for the
Kondo system \cite{schiller03}.

This behavior can be explained by extending the scattering theory
above to include $U$ within the Hartree approximation. This approximation
accounts for $U$ by replacing $e_i\to e_i + U\sum_{j\neq i}\langle
b_j^\dagger b_j\rangle$ where $b_j$ annihilates an electron in the dot state
$e_j$. The charging of the broad resonances can be neglected
as long as we consider chemical potential changes which are small
compared to the bandwidth. With this approximation,
subsequent resonances are separated by $\Delta+U$.

\begin{figure}\centering
\vskip -1.5truecm \epsfxsize6cm\epsfbox{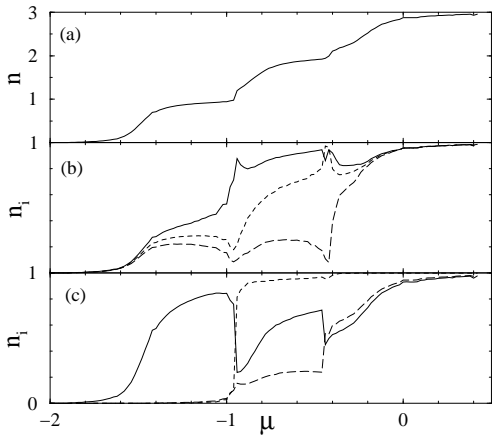}
\vskip .3truecm
\caption{Occupations vs.\ $\mu$
for $N_{\rm dot}=3$ ($\epsilon_1=-1.41t$,$\epsilon_2=-1.4t$, $\epsilon_3=-1.39t$)
coupled to one lead (all couplings $V=0.2t$)
with strong charging energy $U=0.6t$.
(a) Dot occupation $n=n_1+n_2+n_3$. (b) Individual dot-orbital occupations
$n_1$ (full line), $n_2$ (dashed line), $n_3$ (long dashed). (c) Occupation
in the strong-coupling basis: superradiant (full line) and subradiant
states (dashed and long-dashed lines).
} \label{fig3}
\end{figure}

Remarkably, the behavior changes for very large charging energy $U \gg \pi \nu N_{\rm dot}v^2$
where the superradiant state becomes Coulomb blockaded. In this regime, we observe numerically that 
the number of Coulomb-blockade steps equals the number of dot states
$N_{\rm dot}$, cf.\ Fig.\ 3(a). The width of the steps is of order $\pi \nu N_{\rm dot}v^2$, 
large compared to the step widths of subradiant 
states at weaker charging energy. Fig.\ 3(b) shows that the charging steps are due to simultaneous
charging of all three dot orbitals which leads to an oscillatory structure in the 
occupations of the dot states. The origin of these oscillations can be traced by considering the occupations
of the strong-coupling super- and subradiant states (i.e., the eigenstates of $i\pi\nu V^\dagger V$) 
as shown in  
Fig.\ 3(c). Clearly, the origin lies in oscillations in the occupation of the superradiant state
which, in this case, is a symmetric superposition of all dot orbitals.
At the conductance step, one predominantly charges the superradiant state while in between 
steps, there is a tendency to exchange occupations between the superradiant and a 
subradiant state. 

To understand this behavior, consider the configuration of the dot when $\mu$  takes a 
value on the charging plateau $n=1$. In this case, the dot could either charge the superradiant or
a subradiant state. In perturbation theory, the dot-lead coupling changes the (many-body) energy
of these configurations ($E_{\rm super}$ and $E_{\rm sub}$, respectively) due to virtually exciting an 
electron 
from the lead to the superradiant state, if the latter 
is unoccupied, or by virtually exciting an electron from the dot to the leads, if the superradiant state 
is occupied. (We neglect virtual processes involving subradiant states since their coupling to
the leads is much weaker.) This gives an energy difference
$\Delta E=E_{\rm super}-E_{\rm sub} = {W\over2\pi} [\ln (t/|\epsilon+U|)-\ln(t/|\epsilon|)]$ 
between the two configurations, where $\epsilon$ is the energy of the dot state
relative to $\mu$ and $W$ is the width of the superradiant state. (The single-particle level 
spacing is neglected here.) Thus, for $|\epsilon|<U/2$ ($|\epsilon|>U/2$) occupying the superradiant 
(subradiant) state gives the
lower energy so that the charging steps are due to charging of the superradiant state, while somewhere 
on the plateau 
the occupations of the superradiant and subradiant states are exchanged. The precise
location of this switch is affected by the single-particle level spacing of the dot. 
This mechanism was considered by Silvestrov and Imry in a different context \cite{Silvestrov}.

We close with a discussion of possible experimental realizations. 
Generically, for a semiconductor quantum dot perfectly coupled to leads by quantum point 
contacts in the sense that there are $N$ perfectly transmitting channels, one finds that 
$\Gamma\sim N\Delta$, leading to $N^2\Delta/\pi\Gamma\sim N$. This shows that 
this situation is not in the strong coupling limit and thus, there are 
no sharp resonances in agreement with the description of this regime e.g.\ in
Refs.\ \cite{aleiner02,matveev95}.

One situation in which the effect discussed here can be observed is when, by a mesoscopic 
fluctuation or by symmetry, 
several dot levels bunch together so that their effective level spacing is much smaller
than the average $\Delta$. 
Alternatively, a situation with an anomalously small $\Delta$  can be engineered 
by a 
judicious choice of the device. E.g., one can think of a set of $n$ quantum dots with weak interdot
tunneling whose energy levels can be manipulated into almost degeneracy 
by a set of external gates. This realizes a situation with $N_{\rm dot}=n$ and strong coupling 
$\Gamma \gg N^2 \Delta$ to the leads. 
Finally, one may also think of cases in which leads and ``dot'' are made from different materials,
allowing for an independent manipulation of $\Gamma$ and $\Delta$. For example, 
when tunneling 
through a series of identical impurities or a suitable molecule between metallic electrodes, 
one expects $\Gamma$ to be enhanced by the large DOS in the metallic leads. 

Very useful discussions with B.\ Altshuler, I.\ Eremin, Y.\ Gefen, Y.\ Meir, M.\ Raikh,
A. Schiller, P.\ Silvestrov, H.\ Weidenm\"uller, 
and J.\ Weis are gratefully acknowledged, as well 
as support from the Israel Academy of
Science (RB), SFB 290, the ``Junge Akademie" (FvO), and the Minerva Foundation (JK). One
of us (FvO) thanks the Einstein Center at the Weizmann Institute for hospitality while
this work was initiated.

\end{document}